\documentclass
[aps,pra,amsfonts,amssymb,twocolumn,amsmath,preprintnumbers,nofootinbib,floatfix,
showpacs]{revtex4-1}%
\usepackage[dvips]{graphics}
\usepackage{graphicx}
\usepackage{bm}
\usepackage{amsmath}
\usepackage{amsfonts}
\usepackage{amssymb}%
\providecommand{\U}[1]{\protect\rule{.1in}{.1in}}

\begin{document}

\title{Semiclassical Boltzmann theory of spin Hall effects in giant Rashba systems}
\author{Cong Xiao}
\affiliation{Department of Physics, The University of Texas, Austin, Texas 78712-0264, USA}
\begin{abstract}
For the spin Hall effect arising from strong band-structure spin-orbit
coupling, a semiclassical Boltzmann theory reasonably addressing the
intriguing disorder effect called side-jump is still absent. In this paper we
describe such a theory of which the key ingredient is the
spin-current-counterpart of the semiclassical side-jump velocity (introduced
in the context of the anomalous Hall effect). Applying this theory to spin
Hall effects in a two-dimensional electron gas with giant Rashba spin-orbit
coupling, we find largely enhanced spin Hall angle in the presence of magnetic
impurities when only the lower Rashba band is partially occupied.
\end{abstract}
\maketitle

\section{Introduction}

It is now generally accepted that three mechanisms -- intrinsic, side-jump and
skew scattering -- contribute to both the spin Hall effect (SHE) and anomalous
Hall effect (AHE) \cite{Nagaosa2010,Sinova2015}. Among the three mechanisms,
the side-jump mechanism is of special interest because it originates from
scattering but can, in some simple cases
\cite{Nagaosa2010,Sinova2015,Kovalev2010,Culcer2010,Yang2011}, be independent
of both the disorder density and scattering strength. In particular, when the
SHE or AHE arises from strong spin-orbit coupling in the band-structure, the
side-jump belongs to the category of disorder-induced interband-coherence
effect which has recently been an important topic in condensed matter physics
\cite{Nagaosa2010,Sinova2015,Kovalev2010,Sinitsyn,Hou2015,Culcer2017,Xiao2017SOT}%
.

In investigating transport phenomena in solids, the semiclassical Boltzmann
approach is appealing due to its conceptual intuition \cite{Ziman1960}. In the
study of SHE-AHE, how to incorporate side-jump effects into the semiclassical
formalism is an attracting theoretical issue \cite{Nagaosa2010}. In the study
of AHE, the renewed semiclassical theory addressing this issue has proven
useful in obtaining physical pictures \cite{Sinitsyn}. In such a theory, the
quantum mechanical information on side-jump is coded in the expressions of
gauge-invariant classical concepts such as the coordinate-shift and side-jump
velocity \cite{Sinitsyn}. On the other hand, in the field of SHE when the spin
is not conserved due to strong spin-orbit coupling in the band structure, such
as in a Rashba two-dimensional electron gas (2DEG), a semiclassical
description to side-jump SHE is still absent \cite{Zhang2005,noteZhang2005}.
Although the modified Boltzmann equation \cite{Sinitsyn} developed in studying
the AHE can be directly applied to SHE, the spin--current-counterpart of the
side-jump velocity in this case has not been addressed before.

In the present paper we formulate a semiclassical Boltzmann framework of SHE
when the spin is not conserved due to strong band-structure spin-orbit
coupling. This semiclassical theory takes into account interband-coherence
effects induced by both the dc uniform electric field and weak static
disorder. We work out the spin-current-counterpart of the side-jump velocity
based on scattering-induced modifications of conduction-electron states. When
the electric field turns on, this quantity contributes one part of the
side-jump SHE.

As applications we consider the SHE in a 2DEG with giant Rashba spin-orbit
coupling and short-range impurities. We focus on the enhancement of spin Hall
angle when the Fermi energy is tuned down towards and below the band crossing
point in giant Rashba 2DEGs with magnetic disorder. The spin Hall angle which
measures the generation efficiency of the transverse spin current from the
longitudinal electric current is the figure of merit of the SHE. Giant Rashba
spin-orbit coupling energy comparable to or even larger than the Fermi energy
is possible in the polar semiconductor BiTeX (X=Cl, Br and I) family and
related surfaces and interfaces \cite{Eremeev2012,Sakano2013,Wu2014}. Thus
these systems are promising to realize efficient conversion of charge current
into spin current.

The paper is organized as follows. In Sec. II we outline the semiclassical
formulation of SHE. Section III introduces the Rashba model and calculates the
SHE. Section IV concludes the paper.

\section{Semiclassical formulation}

Considering the linear response of the spin current polarized in one
particular direction (z direction is chosen in the following) to a weak dc
uniform electric field $\mathbf{E}$ in non-degenerate multiband electron
systems in the weak disorder regime, one has the semiclassical formula%
\begin{equation}
\mathbf{j}^{z}=\sum_{l}f_{l}\mathbf{j}_{l}^{z}, \label{semi}%
\end{equation}
where\ $\mathbf{j}_{l}^{z}$\ is the amount of spin current carried by the
conduction-electron state denoted by $l$, $f_{l}$ is the semiclassical
distribution function.

The conduction-electron state may be modified by the electric field and static
impurity scattering, $\mathbf{j}_{l}^{z}$ can thus deviate from the customary
pure-band value \cite{Ziman1960,Zhang2005}: $\left(  \mathbf{j}_{l}%
^{z}\right)  ^{0}\equiv\langle l|\mathbf{\hat{\jmath}}^{z}\mathbf{|}l\rangle$
with $\mathbf{\hat{\jmath}}^{z}$ the spin-current operator. Here
$|l\rangle=|\eta\mathbf{k}\rangle=|\mathbf{k}\rangle|u_{\mathbf{k}}^{\eta
}\rangle$ is the Bloch state, $\eta$ is the band index, $\mathbf{k}$ is the
crystal momentum, $|\mathbf{k}\rangle$ and $|u_{\mathbf{k}}^{\eta}\rangle$ are
the plane-wave and periodic parts of $|\eta\mathbf{k}\rangle$, respectively.
Following the recipe based on the quantum-mechanical perturbation theory for
the electric-field modified Bloch state and the Lippmann-Schwinger equation
for the scattering modified conduction-electron state in Ref.
\onlinecite{Xiao2017SOT}, in the weak disorder regime nontrivial corrections
caused by interband-coherence effects to $\left(  \mathbf{j}_{l}^{z}\right)
^{0}$ read:
\begin{equation}
\mathbf{j}_{l}^{z}=\left(  \mathbf{j}_{l}^{z}\right)  ^{0}+\delta
^{in}\mathbf{j}_{l}^{z}+\delta^{ex}\mathbf{j}_{l}^{z}. \label{semi j}%
\end{equation}

The intrinsic correction $\delta^{in}\mathbf{j}_{l}^{z}=2\operatorname{Re}%
\langle l|\mathbf{\hat{\jmath}}^{z}|\delta^{\mathbf{E}}l\rangle$ arises from
the interband-virtual-transition (electron charge e) $|\delta^{\mathbf{E}%
}l\rangle=-i\hbar e\mathbf{E\cdot}\sum_{\eta^{\prime}\neq\eta}|\eta^{\prime
}\mathbf{k}\rangle\langle u_{\mathbf{k}}^{\eta^{\prime}}|\mathbf{\hat{v}%
}|u_{\mathbf{k}}^{\eta}\rangle/\left(  \epsilon_{\mathbf{k}}^{\eta}%
-\epsilon_{\mathbf{k}}^{\eta^{\prime}}\right)  ^{2}$ induced by the electric
field \cite{Zhang2005}, with $\epsilon_{l}\equiv\epsilon_{\mathbf{k}}^{\eta}$
the energy of Bloch state $|\eta\mathbf{k}\rangle$ and $\mathbf{\hat{v}}$\ the
velocity operator. Thus
\begin{equation}
\delta^{in}\mathbf{j}_{l}^{z}=\hbar e\sum_{\eta^{\prime}\neq\eta}%
\frac{2\operatorname{Im}\langle\eta\mathbf{k}|\mathbf{\hat{\jmath}}^{z}%
|\eta^{\prime}\mathbf{k}\rangle\langle u_{\mathbf{k}}^{\eta^{\prime}%
}|\mathbf{\hat{v}\cdot E}|u_{\mathbf{k}}^{\eta}\rangle}{\left(  \epsilon
_{\mathbf{k}}^{\eta}-\epsilon_{\mathbf{k}}^{\eta^{\prime}}\right)  ^{2}}%
\end{equation}
is an electric-field-induced interband-coherence effect.

The extrinsic correction $\delta^{ex}\mathbf{j}_{l}^{z}$ originates from the
interband-coherence during the elastic electron-impurity scattering
process.\ The scattering-induced modification to conduction-electron states is
captured by the Lippmann-Schwinger equation describing the scattering state
$|l^{s}\rangle=|l\rangle+\left(  \epsilon_{l}-\hat{H}_{0}+i\epsilon\right)
^{-1}\hat{T}|l\rangle$ with the T-matrix $\hat{T}|l\rangle=\hat{V}%
|l^{s}\rangle$ related to the disorder potential $\hat{V}$ and disorder-free
Hamiltonian $\hat{H}_{0}$. $|\delta l^{s}\rangle\equiv|l^{s}\rangle-|l\rangle$
denotes the scattering-induced modification to the Bloch state. Thus
$\delta^{ex}\mathbf{j}_{l}^{z}$ is related to the values of
$2\operatorname{Re}\left\langle \langle l|\mathbf{\hat{\jmath}}^{z}|\delta
l^{s}\rangle\right\rangle _{c}$ and $\left\langle \langle\delta l^{s}%
|\mathbf{\hat{\jmath}}^{z}|\delta l^{s}\rangle\right\rangle _{c}$\ in the
lowest nonzero order in the disorder potential. Here $\left\langle
..\right\rangle _{c}$ denotes the average over disorder configurations and we
assume that the statistical average of the disorder potential is zero (nonzero
value only shifts the origin of total energy) $\left\langle V\right\rangle
_{c}=0$. Only the terms containing interband matrix elements of $\mathbf{\hat
{\jmath}}^{z}$ represent the disorder-induced interband-coherence effects,
therefore \cite{note-SO scattering}
\begin{align}
\delta^{ex}\mathbf{j}_{l}^{z}  &  =\sum_{\eta^{\prime\prime}\neq\eta^{\prime}%
}\sum_{\eta^{\prime}\mathbf{k}^{\prime}}\nonumber\\
&  \times\frac{\left\langle \langle\eta\mathbf{k}|\hat{V}\mathbf{|}%
\eta^{\prime}\mathbf{k}^{\prime}\rangle\langle\eta^{\prime\prime}%
\mathbf{k}^{\prime}|\hat{V}\mathbf{|}\eta\mathbf{k}\rangle\right\rangle
_{c}\langle\eta^{\prime}\mathbf{k}^{\prime}|\mathbf{\hat{\jmath}}%
^{z}\mathbf{|}\eta^{\prime\prime}\mathbf{k}^{\prime}\rangle}{\left(
\epsilon_{\mathbf{k}}^{\eta}-\epsilon_{\mathbf{k}^{\prime}}^{\eta^{\prime}%
}-i0^{+}\right)  \left(  \epsilon_{\mathbf{k}}^{\eta}-\epsilon_{\mathbf{k}%
^{\prime}}^{\eta^{\prime\prime}}+i0^{+}\right)  }\nonumber\\
&  +2\operatorname{Re}\sum_{\eta^{\prime}\neq\eta}\sum_{\eta^{\prime\prime
}\mathbf{k}^{\prime\prime}}\nonumber\\
&  \times\frac{\left\langle \langle\eta^{\prime}\mathbf{k}|\hat{V}%
\mathbf{|}\eta^{\prime\prime}\mathbf{k}^{\prime\prime}\rangle\langle
\eta^{\prime\prime}\mathbf{k}^{\prime\prime}|\hat{V}\mathbf{|}\eta
\mathbf{k}\rangle\right\rangle _{c}\langle\eta\mathbf{k}|\mathbf{\hat{\jmath}%
}^{z}\mathbf{|}\eta^{\prime}\mathbf{k}\rangle}{\left(  \epsilon_{\mathbf{k}%
}^{\eta}-\epsilon_{\mathbf{k}}^{\eta^{\prime}}+i0^{+}\right)  \left(
\epsilon_{\mathbf{k}}^{\eta}-\epsilon_{\mathbf{k}^{\prime\prime}}%
^{\eta^{\prime\prime}}+i0^{+}\right)  }. \label{ex}%
\end{align}
It has been shown \cite{Xiao2017SOT} that the side-jump velocity
$\mathbf{v}_{l}^{sj}$ which is an important ingredient in the semiclassical
theory of AHE \cite{Sinitsyn} can also be obtained in this way ($\delta
^{ex}\mathbf{v}_{l}=\mathbf{v}_{l}^{sj}$) and thus shares the same origin.
$\delta^{ex}\mathbf{j}_{l}^{z}$ can therefore be deemed as the
spin--current-counterpart of the side-jump velocity in the case of
band-structure spin-orbit coupling.

The properly modified steady-state linearized Boltzmann equation in the
presence of weak static disorder has been proposed as \cite{Sinitsyn}:%
\begin{equation}
e\mathbf{E}\cdot\mathbf{v}_{l}^{0}\frac{\partial f^{0}}{\partial\epsilon_{l}%
}=-\sum_{l^{\prime}}\omega_{l,l^{\prime}}\left[  f_{l}-f_{l^{\prime}}%
-\frac{\partial f^{0}}{\partial\epsilon_{l}}e\mathbf{E}\cdot\delta
\mathbf{r}_{l\prime,l}\right]  ,\label{SBE}%
\end{equation}
where $\mathbf{v}_{l}^{0}=\partial\epsilon_{l}/\hbar\partial\mathbf{k}$ is the
band velocity, $f^{0}$ is the Fermi distribution function, $\delta
\mathbf{r}_{l\prime,l}$ is the coordinate-shift in the scattering process
($l\rightarrow l^{\prime}$) \cite{Sinitsyn} and $\omega_{l,l^{\prime}}$ the
semiclassical scattering rate ($l^{\prime}\rightarrow l$). Up to the linear
order of the electric field\ one has the decomposition
\cite{Sinitsyn,Xiao2017AHE}
\begin{equation}
f_{l}=f_{l}^{0}+g_{l}^{n}+g_{l}^{a},
\end{equation}
with $g_{l}^{n}$ the normal part of the out-of-equilibrium distribution
function satisfying the Boltzmann equation in the absence of $\delta
\mathbf{r}_{l\prime,l}$ and $g_{l}^{a}$ the anomalous distribution function
related to $\delta\mathbf{r}_{l\prime,l}$. It is now clear \cite{Sinitsyn}
that $\delta r_{l\prime,l}$ is a disorder-induced interband-coherence effect
and so is $g_{l}^{a}$.

Given that the semiclassical formulation is relevant in the weak disorder
regime, Eq. (\ref{semi}) reduces to \cite{Xiao2017SOT}
\begin{equation}
\mathbf{j}^{z}=\sum_{l}f_{l}\left(  \mathbf{j}_{l}^{z}\right)  ^{0}+\sum
_{l}g_{l}^{2s}\delta^{ex}\mathbf{j}_{l}^{z}+\sum_{l}f_{l}^{0}\delta
^{in}\mathbf{j}_{l}^{z}, \label{SBE response}%
\end{equation}
up to the zeroth order of total impurity density and scattering strength.
$g_{l}^{2s}$ represents the value of $g_{l}^{n}$ in the lowest Born order
\cite{Sinitsyn}. In higher Born orders, some additional contributions to
$g_{l}^{n}$ appear and are responsible for the transverse transport due to the
breakdown of the principle of microscopic detailed balance. The analysis of
these higher-Born-order contributions under the non-crossing approximation has
been detailed in Ref. \onlinecite{Xiao2017AHE}. Here we only mention that
there is an interband-coherence scattering effect called \textquotedblleft
intrinsic-skew-scattering induced side-jump\textquotedblright\ appearing in
the third Born order under the Gaussian disorder. Below we set $\mathbf{j}%
^{z,in}=\sum_{l}f_{l}^{0}\delta^{in}\mathbf{j}_{l}^{z}$ which is just the
intrinsic contribution to the spin current independent of the disorder
\cite{Zhang2005}, and $\mathbf{j}^{z,sj}=\sum_{l}g_{l}^{2s}\delta
^{ex}\mathbf{j}_{l}^{z}$ because it is related to the
spin--current-counterpart of the side-jump velocity. In general case of SHE
induced by strong band-structure spin-orbit coupling, $\mathbf{j}^{z,sj}$ is
just one part of the side-jump SHE arising from disorder-induced
interband-coherence effects. Other two semiclassical contributions to the
side-jump SHE (from the anomalous distribution function $g_{l}^{a}$ and the
intrinsic-skew-scattering induced side-jump) \cite{note-sj} and the skew
scattering SHE arising from non-Gaussian disorder are all included in the
first term of Eq. (\ref{SBE response}) \cite{Sinitsyn,Xiao2017AHE}.

To be more clear we can consider the case of randomly distributed scalar
pointlike Gaussian disorder with density $n_{0}$ and average strength $V_{0}$.
Then $g_{l}^{2s}\sim n_{0}^{-1}V_{0}^{-2}$, $\delta^{ex}j_{l}^{z}\sim
n_{0}V_{0}^{2}$, $g_{l}^{a}\sim n_{0}^{0}V_{0}^{0}$, and the third-Born-order
contribution to $g_{l}^{n}$ behaves as $\sim n_{0}^{0}V_{0}^{0}$ (thus is
called the intrinsic-skew-scattering \cite{Sinitsyn,Xiao2017AHE}). In this
case the side-jump SHE may consist of three semiclassical contributions in the
zeroth order of both the impurity density and scattering strength: $j^{z,sj}$,
$\sum_{l}g_{l}^{a}\left(  \mathbf{j}_{l}^{z}\right)  ^{0}$ and the
intrinsic-skew-scattering induced side-jump.

\section{Model calculation}

\subsection{Model}

The model Hamiltonian of a Rashba 2DEG is $H_{0}=\frac{\hbar^{2}\mathbf{k}%
^{2}}{2m}+\alpha_{R}\mathbf{\hat{\sigma}}\cdot\left(  \mathbf{k}%
\times\mathbf{\hat{z}}\right)  $, where $\mathbf{k}$ is the 2D wavevector, $m$
is the effective mass, $\mathbf{\hat{\sigma}}$ is the vector of Pauli
matrices, $\alpha_{R}$ the Rashba coefficient. The internal eigenstates read
$|u_{\mathbf{k}}^{\eta}\rangle=\frac{1}{\sqrt{2}}\left[  1,-i\eta\exp\left(
i\phi\right)  \right]  ^{T}$, where $\eta=\pm$ label the two bands
$\epsilon_{k}^{\eta}=\hbar^{2}k^{2}/\left(  2m\right)  +\eta\alpha_{R}k$, and
$\tan\phi=k_{y}/k_{x}$.

For $\epsilon>0$ the corresponding wave number in $\eta$ band is given as
$k_{\eta}\left(  \epsilon\right)  =-\eta k_{R}+k_{0}\left(  \epsilon\right)
$. Here $k_{R}=m\alpha_{R}/\hbar^{2}=\frac{1}{2}\left(  k_{-}\left(
\epsilon\right)  -k_{+}\left(  \epsilon\right)  \right)  $ measures the
momentum splitting of two Rashba bands, whereas $k_{0}\left(  \epsilon\right)
\equiv\alpha_{R}^{-1}\sqrt{\epsilon_{R}^{2}+2\epsilon_{R}\epsilon}=\frac{1}%
{2}\sum_{\eta}k_{\eta}\left(  \epsilon\right)  $. The density of states of
$\eta$ band takes the form $D_{\eta}\left(  \epsilon\right)  =D_{0}%
\frac{k_{\eta}\left(  \epsilon\right)  }{k_{0}\left(  \epsilon\right)  }$,
with $D_{0}=m/2\pi\hbar^{2}$.

For $0>\epsilon>-\epsilon_{R}/2$, the iso-energy surface slices the spectrum
two rings of radii $k_{-\nu}\left(  \epsilon\right)  =k_{R}+\left(  -1\right)
^{\nu-1}k_{0}\left(  \epsilon\right)  $, where $\nu=1,2$ denote the two
monotonic segments (Fig. 1), $k_{0}\left(  \epsilon\right)  \equiv\alpha
_{R}^{-1}\sqrt{\epsilon_{R}^{2}+2\epsilon_{R}\epsilon}=\frac{1}{2}\left(
k_{-1}\left(  \epsilon\right)  -k_{-2}\left(  \epsilon\right)  \right)  $. The
density of states of $-\nu$ branch reads $D_{-\nu}\left(  \epsilon\right)
=D_{0}\frac{k_{-\nu}\left(  \epsilon\right)  }{k_{0}\left(  \epsilon\right)
}$.\begin{figure}[ptb]
\includegraphics[width=0.35\textwidth]{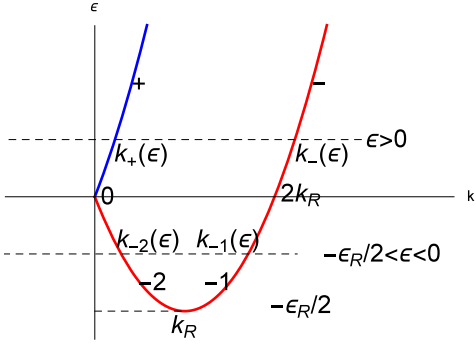} \caption{Band
structure of Rashba 2DEG. The two Rashba bands cross at the zero-momentum
point. The minimal energy of the dispersion curve is $-\frac{1}{2}\epsilon
_{R}$. For $\epsilon\geq0$, the wave number in $\pm$ band is defined by
$k_{\pm}\left(  \epsilon\right)  $. For $-\frac{1}{2}\epsilon_{R}<\epsilon
\leq0$, there exist two monotonic branches: the one from $k=k_{R}$ to $2k_{R}$
is marked by the branch $-1$, the other from $k=0$ to $k_{R}$ marked by branch
$-2$. $k_{-\nu}\left(  \epsilon\right)  $ denotes the wave number in the
$-\nu$ branch at given $\epsilon$, where $\nu=1,2$.}%
\label{fig1}%
\end{figure}

The conventional definition of the spin current as an anti-commutator of
velocity and spin is employed: $\mathbf{\hat{\jmath}}^{z}=\frac{\hbar}{2}%
\frac{1}{2}\left\{  \hat{\sigma}_{z},\mathbf{\hat{v}}\right\}  =\frac{\hbar
}{2}\frac{\hbar\mathbf{k}}{m}\hat{\sigma}_{z}$. It is purely off-diagonal in
band-index space in this model: $\left(  \mathbf{j}_{l}^{z}\right)  ^{0}=0$,
thus the SHE in Eq. (\ref{SBE response}) is determined only by%
\begin{equation}
\mathbf{j}^{z}=\mathbf{j}^{z,in}+\mathbf{j}^{z,sj}.
\end{equation}

The Boltzmann equation can be conveniently solved by using variables
$l=\left(  \epsilon,\eta,\phi\right)  $ for $\epsilon>0$ and $l=\left(
\epsilon,-\nu,\phi\right)  $ for $0>\epsilon>-\epsilon_{R}/2$.
Correspondingly, $\sum_{l}=\sum_{\eta\left(  \nu\right)  }\int d\epsilon
D_{\eta\left(  -\nu\right)  }\left(  \epsilon\right)  \int\frac{d\phi}{2\pi}$
for $\epsilon>0$ $\left(  0>\epsilon>-\epsilon_{R}/2\right)  $. If
$\epsilon>0$, in the lowest Born order the energy-integrated elastic
scattering rate is $\omega_{\eta,\eta^{\prime}}^{\phi,\phi^{\prime}}\left(
\epsilon\right)  =\int d\epsilon^{\prime}D_{\eta^{\prime}}\left(
\epsilon^{\prime}\right)  \omega_{l\prime,l}^{2s}$. Whereas if $0>\epsilon
>-\epsilon_{R}/2$ there exists elastic scattering between the two branches
$\nu=1$ and $2$, and one has $\omega_{-\nu,-\nu^{\prime}}^{\phi,\phi^{\prime}%
}\left(  \epsilon\right)  =D_{-\nu^{\prime}}\left(  \epsilon\right)  \int
d\epsilon_{l^{\prime}}\omega_{l\prime,l}^{2s}$. Assuming isotropic disorder
potential, transport-time type solutions to $g_{l}^{2s}$ exist
\cite{Xiao2016PRB}. For $\epsilon>0$ we have%
\begin{equation}
g_{\eta}^{2s}\left(  \epsilon\right)  =\left(  -\partial_{\epsilon}%
f^{0}\right)  e\mathbf{E}\cdot\mathbf{v}_{\eta}\left(  \epsilon,\phi\right)
\tau_{\eta}^{tr}\left(  \epsilon\right)  ,
\end{equation}
where the transport time $\tau_{\eta}^{tr}\left(  \epsilon\right)  $ is
determined by
\begin{equation}
1=\sum_{\eta^{\prime}}\int\frac{d\phi^{\prime}}{2\pi}\omega_{\eta,\eta
^{\prime}}^{\phi,\phi^{\prime}}\left[  \tau_{\eta}^{tr}-\cos\left(
\phi^{\prime}-\phi\right)  \tau_{\eta^{\prime}}^{tr}\right]  .
\end{equation}
For $0>\epsilon>-\epsilon_{R}/2$ we have%
\begin{equation}
g_{-\nu}^{2s}\left(  \epsilon\right)  =\left(  -\partial_{\epsilon}%
f^{0}\right)  e\mathbf{E}\cdot\mathbf{v}_{-\nu}\left(  \epsilon,\phi\right)
\tau_{-\nu}^{tr}\left(  \epsilon\right)  ,
\end{equation}
with the transport time $\tau_{-\nu}^{tr}\left(  \epsilon\right)  $ decided
by
\begin{equation}
1=\sum_{\nu^{\prime}}\int\frac{d\phi^{\prime}}{2\pi}\omega_{-\nu,-\nu^{\prime
}}^{\phi,\phi^{\prime}}\left[  \tau_{-\nu}^{tr}-\left(  -1\right)
^{\nu^{\prime}-\nu}\cos\left(  \phi^{\prime}-\phi\right)  \tau_{-\nu^{\prime}%
}^{tr}\right]  .
\end{equation}

\subsection{Calculations}

We consider the impurity potential is produced by randomly distributed
short-range scatters at $\mathbf{R}_{i}$, i.e., $V\left(  \mathbf{r}\right)
=\sum_{i,\mu}V_{\mu}^{i}\sigma_{\mu}\delta\left(  \mathbf{r}-\mathbf{R}%
_{i}\right)  $ with $\mu=0,1,2,3$ and $\sigma_{0}$ the unity matrix in spin
space \cite{Lu2013}. Here the short-range potential is approximated by the
delta-potential. We assume Gaussian disorder approximation and isotropic
magnetic scattering \cite{Lu2013,Inoue2006}. $n_{0}$ and $n_{m}$ are the
concentrations of nonmagnetic and magnetic impurities, respectively. $V_{0}$
and $V_{m}$ are the average strengths for the nonmagnetic and magnetic
scattering, respectively. The external electric field is applied in x direction.

\subsubsection{Nonmagnetic impurities}

When $\epsilon>0$, straightforward calculation leads to the
spin-current-counterpart of the side-jump velocity
\begin{equation}
\delta^{ex}\left(  \mathbf{j}_{l}^{z}\right)  _{y}^{nm}=-\frac{1}{\tau_{0}%
}\frac{\eta}{k_{R}}\frac{\hbar}{8}\cos\phi, \label{sj-nm-above}%
\end{equation}
with $\tau_{0}=\left(  2\pi n_{0}V_{0}^{2}D_{0}/\hbar\right)  ^{-1}$. The
transport time reads \cite{Xiao2016PRB} $\tau_{\eta}^{tr}\left(
\epsilon\right)  =\tau_{0}D_{\eta}\left(  \epsilon\right)  /D_{0}$, and then
the side-jump spin Hall current is%
\begin{equation}
j_{y}^{z,sj}=\sum_{l}g_{l}^{2s}\delta^{ex}\left(  \mathbf{j}_{l}^{z}\right)
_{y}^{nm}=\frac{e}{8\pi}E_{x},
\end{equation}
which completely cancels out the intrinsic spin Hall current $j_{y}%
^{z,in}=\frac{-e}{8\pi}E_{x}$. This just reproduces the well-known
\cite{Sinova2015} vanishing spin Hall current $j_{y}^{z}=0$ in the
semiclassical Boltzmann theory for the first time.

When $0>\epsilon_{F}>-\epsilon_{R}/2$, the intrinsic spin Hall current is
$j_{y}^{z,in}=\frac{k_{0}\left(  \epsilon_{F}\right)  }{k_{R}}\frac{-e}{8\pi
}E_{x}$. Meanwhile the spin-current-counterpart of the side-jump velocity
reads%
\begin{equation}
\delta^{ex}\left(  \mathbf{j}_{l}^{z}\right)  _{y}^{nm}=\frac{1}{\tau_{0}%
}\frac{1}{k_{0}\left(  \epsilon\right)  }\frac{\hbar}{8}\cos\phi.
\label{sj-nm-below}%
\end{equation}
and thus the side-jump spin Hall current%
\begin{equation}
j_{y}^{z,sj}=\sum_{l}g_{l}^{2s}\delta^{ex}\left(  \mathbf{j}_{l}^{z}\right)
_{y}^{nm}=\frac{k_{0}\left(  \epsilon_{F}\right)  }{k_{R}}\frac{e}{8\pi}E_{x}%
\end{equation}
again cancels out the intrinsic one. This also coincides with the zero SHE
obtained by the Kubo formula \cite{Grimaldi2006}.

\subsubsection{Magnetic impurities}

For isotropic delta-like magnetic impurity potential, since the contributions
from $V_{x}^{i}$ and $V_{y}^{i}$ cancel out in Eq. (\ref{ex}), the
spin--current-counterpart of the side-jump velocity is given by%
\begin{equation}
\delta^{ex}\left(  \mathbf{j}_{l}^{z}\right)  _{y}^{m}=-\frac{1}{3}\frac
{n_{m}V_{m}^{2}}{n_{0}V_{0}^{2}}\delta^{ex}\left(  \mathbf{j}_{l}^{z}\right)
_{y}^{nm}.
\end{equation}
The transport time is given by
\begin{equation}
\frac{\tau_{\eta}^{tr}\left(  \epsilon\right)  }{\tau_{m}}=\frac{8k_{0}\left(
\epsilon\right)  -k_{\eta}\left(  \epsilon\right)  }{7k_{0}\left(
\epsilon\right)  }%
\end{equation}
for $\epsilon>0$, and
\begin{equation}
\frac{\tau_{-\nu}^{tr}\left(  \epsilon\right)  }{\tau_{m}}=\frac{k_{0}\left(
\epsilon\right)  }{k_{R}}\frac{8k_{R}-k_{-\nu}\left(  \epsilon\right)
}{7k_{R}}%
\end{equation}
for $0>\epsilon>-\epsilon_{R}/2$, with $\tau_{m}=\left(  \frac{2\pi}{\hbar
}n_{m}V_{m}^{2}D_{0}\right)  ^{-1}$.

When both Rashba bands are partially occupied, the side-jump spin Hall current%
\begin{equation}
j_{y}^{z,sj}=\sum_{l}g_{l}^{2s}\delta^{ex}\left(  \mathbf{j}_{l}^{z}\right)
_{y}^{m}=\frac{1}{7}\frac{-e}{8\pi}E_{x}=\frac{1}{7}j_{y}^{z,in}%
\end{equation}
enhances the total spin Hall current to $j_{y}^{z}=j_{y}^{z,in}+j_{y}%
^{z,sj}=\frac{8}{7}j_{y}^{z,in}$. This $j_{y}^{z}$ is the same as the
weak-disorder-limit value of that obtained by Kubo diagrammatic calculations
\cite{Inoue2006,Wang2007}. The longitudinal electric current is $j_{x}%
=\frac{e^{2}}{\pi\hbar^{2}}\tau_{m}\frac{3\epsilon_{R}+7\epsilon_{F}}{7}E_{x}%
$, the spin Hall angle is therefore%
\begin{equation}
\alpha_{sH}=\frac{ej_{y}^{z}/\left(  \frac{\hbar}{2}\right)  }{j_{x}}%
=\frac{-2\hbar}{\tau_{m}\epsilon_{R}}\frac{1}{3+7\frac{\epsilon_{F}}%
{\epsilon_{R}}}.
\end{equation}

When only the lower Rashba band is partially occupied, the side-jump and the
total spin Hall currents are%
\begin{equation}
j_{y}^{z,sj}=\frac{k_{0}\left(  \epsilon_{F}\right)  }{7k_{R}}\frac{-e}{8\pi
}E_{x}=\frac{1}{7}j_{y}^{z,in}%
\end{equation}
and $j_{y}^{z}=\frac{8}{7}j_{y}^{z,in}$, respectively. The longitudinal
electric current is $j_{x}=\frac{e^{2}}{\pi\hbar^{2}}\tau_{m}\frac
{3\epsilon_{R}-\epsilon_{F}}{7}\frac{k_{0}^{2}\left(  \epsilon_{F}\right)
}{k_{R}^{2}}E_{x}$, thus
\begin{equation}
\alpha_{sH}=\frac{-2\hbar}{\tau_{m}\epsilon_{R}}\frac{1}{\left(
3-\frac{\epsilon_{F}}{\epsilon_{R}}\right)  \sqrt{1+2\frac{\epsilon_{F}%
}{\epsilon_{R}}}}.
\end{equation}
Although $\frac{\hbar}{\tau_{m}\epsilon_{R}}$ is a small quantity in giant
Rashba systems, the factor $\sqrt{1+2\frac{\epsilon_{F}}{\epsilon_{R}}}$ can
be very small leading to a large spin Hall angle when $\epsilon_{F}$ is
located close to the band bottom of the lower Rashba band. For instance, if
$\frac{\hbar}{\tau_{m}\epsilon_{R}}=0.02$, $1+2\frac{\epsilon_{F}}%
{\epsilon_{R}}=0.1$ leads to $\alpha_{sH}\simeq-4\%$, which is quite large
\cite{Ebert2015,Fert2011}. Smaller $\tau_{m}$ and smaller $\sqrt
{1+2\frac{\epsilon_{F}}{\epsilon_{R}}}$\ may lead to larger $\alpha_{sH}$.
However, the quantitative analysis of this possibility is beyond the scope of
the semiclassical theory which is valid only in the weak disorder regime. From
the above equation, $\alpha_{sH}$ goes to infinity as $\epsilon_{F}$ goes to
the band bottom of the lower Rashba band. But this low carrier density limit
is actually beyond the Boltzmann regime, and more rigorous microscopic
treatments are called for.

\subsubsection{Both nonmagnetic and magnetic impurities}

The coexistence of nonmagnetic and magnetic impurities may be the more
realistic case \cite{Inoue2006,Lu2013}. Only main results will be given in
this case. Since there is no mixing between the nonmagnetic and magnetic
scattering as pointed out by Inoue et al. \cite{Inoue2006}, the
spin-current-counterpart of the side-jump velocity is $\delta^{ex}\left(
\mathbf{j}_{l}^{z}\right)  _{y}=\left[  1-\frac{1}{3}\frac{\tau_{0}}{\tau_{m}%
}\right]  \delta^{ex}\left(  \mathbf{j}_{l}^{z}\right)  _{y}^{nm}$. The total
spin Hall current reads $j_{y}^{z}=\frac{\frac{8}{3}\frac{\tau_{0}}{\tau_{m}}%
}{1+\frac{7}{3}\frac{\tau_{0}}{\tau_{m}}}j_{y}^{z,in}$, depending on the
relative weight of different types of scattering \cite{Yang2011,Lu2013}. The
side-jump effect vanishes when $\tau_{0}=3\tau_{m}$, the same condition as
that for the vanishing of the ladder vertex correction in the Kubo
diagrammatic calculation in the spin-$\sigma_{z}$ basis for the case
$\epsilon_{F}>0$ \cite{Inoue2006,note-vertex}.

For Fermi energies above and below the band crossing point, the spin Hall
angles are
\[
\alpha_{sH}=\frac{-\frac{\hbar}{\tau_{S}\epsilon_{R}}\frac{2}{3}\frac{\tau
_{0}}{\tau_{m}}}{1+\frac{\tau_{0}}{\tau_{m}}+\left(  1+\frac{7}{3}\frac
{\tau_{0}}{\tau_{m}}\right)  \frac{\epsilon_{F}}{\epsilon_{R}}}%
\]
and
\[
\alpha_{sH}=\frac{-\frac{\hbar}{\tau_{S}\epsilon_{R}}\frac{2}{3}\frac{\tau
_{0}}{\tau_{m}}}{1+\frac{\tau_{0}}{\tau_{m}}+\left[  1-\frac{1}{3}\frac
{\tau_{0}}{\tau_{m}}\right]  \frac{\epsilon_{F}}{\epsilon_{R}}}\frac{1}%
{\sqrt{1+2\frac{\epsilon_{F}}{\epsilon_{R}}}},
\]
respectively. Here we define $\tau_{S}^{-1}\equiv\tau_{0}^{-1}+\tau_{m}^{-1}$.
Tuning the ratio $\tau_{0}/\tau_{m}$\ one can find that $\alpha_{sH}$ changes
monotonically and continuously from the scalar-disorder-dominated case to the
magnetic-disorder-dominated regime.

\section{Discussion and Summary}

Before concluding this paper, we comment on some important issues not
mentioned in above sections.

First, the simple form of the semiclassical Boltzmann equation (\ref{SBE}) is
exactly valid only for isotropic bands and isotropic scattering
\cite{Nagaosa2010,Xiao2017AHE}. In our opinion, in the presence of anisotropy
a more generic and complicated form of the Boltzmann equation may be
necessary, we refer the readers to Ref. \onlinecite{Luttinger1957} for
detailed discussions.

Second, the recently highlighted \textquotedblleft coherent skew
scattering\textquotedblright\ under the Gaussian disorder beyond the
non-crossing approximation \cite{Ado2015} is also included in the first term
of Eq. (\ref{SBE response}). This additional contribution is also in the
zeroth order of both the impurity density and scattering strength in the weak
disorder limit in the presence of only one type of disorder, like the
side-jump contribution, but is not an interband-coherence scattering effect
\cite{note-sj}. Thus how to place this contribution into the classification of
AHE-SHE mechanisms suggested in Refs. \onlinecite{Nagaosa2010,Sinova2015} is
still an open question. Therefore, in presenting our theory we avoid this
issue. Fortunately, in the Rashba model considered in Sec. III the first term
of Eq. (\ref{SBE response}) vanishes. Besides, we should remind the interested
readers that this so-called \textquotedblleft coherent skew
scattering\textquotedblright\ has actually already been proposed sixty years
ago by Kohn and Luttinger \cite{Luttinger1957,Luttinger1958}. We will provide
a comprehensive description of a semiclassical Boltzmann theory going beyond
the non-crossing approximation in a future publication.

Finally, in the presence of spin-orbit coupling the electron spin is not
conserved thus the spin current is not uniquely defined. The conventionally
defined spin current adopted in this study is not a conserved transport
current. A physically attracting definition of the conserved spin current has
been suggested by Shi et al. \cite{Shi2006} by introducing the torque dipole
moment. However, disorder effects on the torque dipole spin current
\cite{Nagaosa2006} in the Bloch representation are hard to deal with under the
uniform external electric field in the Boltzmann theory. We reserve these for
future studies.

In summary, we have formulated a semiclassical Boltzmann framework of spin
Hall effects induced by strong band-structure spin-orbit coupling in
non-degenerate multiband electron systems in the weak disorder regime. We
worked out the absent ingredient in previous semiclassical theories, i.e., the
spin--current-counterpart of the semiclassical side-jump velocity. This
gauge-invariant quantity arises from the interband-coherence during the
elastic electron-impurity scattering, and contributes one part of the
side-jump spin Hall effect.

Applying this theory to a 2DEG with giant Rashba spin-orbit coupling, we
showed an enhanced spin Hall angle when only the lower Rashba band is
partially occupied in the presence of magnetic impurities. We note that this
energy regime below the band crossing point in Rashba systems and similar
systems is of intense theoretical interest also from the standpoint of
enhanced efficiency of spin-orbit torque and of Edelstein effect
\cite{Murakami2012,Xiao2016FOP,Zhang2016}, as well as enhanced thermoelectric
conversion efficiency \cite{Wu2014,Xiao2016PRB}.

\begin{acknowledgments}
The author is indebted to Qian Niu, Dingping Li and Zhongshui Ma for insightful discussions.
\end{acknowledgments}

\end{document}